\documentclass{article} 
\usepackage{lprocl}
\usepackage{axodraw}
\usepackage{rotating}
\newcommand{\fd}{f_D}
\newcommand{\fds}{f_{D_s}}
\newcommand{\fb}{f_B}
\newcommand{\fbd}{f_{B_d}}
\newcommand{\fbs}{f_{B_s}}
\newcommand{\bb}{B_B}

\newcommand{\rgbb}{\hat{B}_B}
\newcommand{\rgbbd}{\hat{B}_{B_d}}
\newcommand{\rgbbs}{\hat{B}_{B_s}}

\newcommand{\MSbar}{\overline{\mbox{\small MS}}}
\def\btorho{\bar B^0\to\rho^+ l^- \bar\nu_l}
\def\btopi{\bar B^0\to\pi^+ l^- \bar\nu_l}
\def\btokstargamma{\bar B\to K^* \gamma}

\def\qsqmax{q^2_{\mathrm{max}}}
\def\w{\omega}

\def\etal{et al.}
\def\kev{\,\mathrm{ke\kern-0.1em V}}
\def\mev{\,\mathrm{Me\kern-0.1em V}}
\def\gev{\,\mathrm{Ge\kern-0.1em V}}
\def\tev{\,\mathrm{Te\kern-0.1em V}}
\def\gtrsim{\mathrel{\lower .7ex\hbox{$\buildrel\textstyle>\over\sim$}}}
\def\lesssim{\mathrel{\lower .7ex\hbox{$\buildrel\textstyle<\over\sim$}}}
\def\dotprod{\mathord{\cdot}}
\def\vec#1{\mathbf{#1}}

\newcommand{\csw}{c_\mathrm{sw}}

\newcommand{\err}[2]{%
{{\renewcommand{\arraystretch}{0.4}%
\ensuremath{\mathop{\raisebox{0.1\height}{\scriptsize
$\begin{array}{@{}c@{}}+\\-\end{array}$}}%
\raisebox{0.1\height}{\scriptsize
$\begin{array}{@{}r@{}}#1\\#2\end{array}$}}}}%
}

\let\errp\errparen
\def\dotprod{\mathord{\cdot}}
\def\vec#1{\mathbf{#1}}

\newdimen\unit
\def\point#1 #2 #3{\vbox to0pt{\kern-#2\unit
  \hbox{\kern#1\unit$#3$}\vss}
 \nointerlineskip}

\def\be{\begin{equation}}
\def\ee{\end{equation}}
\def\bea{\begin{eqnarray}}
\def\eea{\end{eqnarray}}

\begin{document}
\begin{flushright}SHEP 97/26\\
hep-ph/9711386
\end{flushright}
\vspace{3em}

\begin{center}
{\Large\bfseries Theoretical Aspects of Heavy Flavour Physics}\\[2em]
C.T.~Sachrajda\\[1em]
Department of Physics and Astronomy, University of Southampton\\
Southampton SO17 1BJ, UK
\end{center}
\vfill

\begin{center}\textbf{Abstract}\end{center}
\begin{quote}
I review the status of theoretical aspects of
$B$-decays. The principal difficulty in interpreting the wealth of
experimental data is the control of non-perturbative QCD effects, and
the talk is focused on attempts to control these effects. Lattice
results for the decay constants, $B$-$\bar B$ mixing and semileptonic
form-factors are summarized. The discrepancy of the theoretical
predictions and experimental measurements for the ratio of lifetimes
$\tau(\Lambda_b)/\tau(B_0)$ is discussed, as well as the status of the
semileptonic branching ratio of the $B$-meson. The difficulties in
making quantitative predictions for exclusive nonleptonic decays are
stressed, and some recent approaches to this problem are outlined.
\end{quote}
\vfill

\begin{center}
To appear in the proceedings of the XVIII International Symposium
on Lepton-Photon Interactions, Hamburg, Germany, July 28-August 1st 1997.
\end{center}
\vfill

\begin{flushleft}
October 1997
\end{flushleft}

\newpage
\pagestyle{plain}
\setcounter{page}{1}
\title{THEORETICAL ASPECTS OF HEAVY FLAVOUR PHYSICS}

\author{C.~T.~Sachrajda}

\address{Department of Physics and Astronomy,\\ 
University of Southampton,\\ 
Southampton, SO17 1BJ, UK}

\maketitle\abstracts{I review the status of theoretical aspects of
$B$-decays. The principal difficulty in interpreting the wealth of
experimental data is the control of non-perturbative QCD effects, and
the talk is focused on attempts to control these effects. Lattice
results for the decay constants, $B$-$\bar B$ mixing and semileptonic
form-factors are summarized. The discrepancy of the theoretical
predictions and experimental measurements for the ratio of lifetimes
$\tau(\Lambda_b)/\tau(B_0)$ is discussed, as well as the status of the
semileptonic branching ratio of the $B$-meson. The difficulties in
making quantitative predictions for exclusive nonleptonic decays are
stressed, and some recent approaches to this problem are outlined.}

\section{Introduction}
\label{sec:intro}

Weak decays of heavy quarks are a particularly fertile field of
research for detailed tests of the Standard Model of Particle Physics,
for measurements of its parameters (the Cabibbo-Kobayashi-Maskawa
(CKM) matrix elements in particular) and for potential signatures of
{\em new physics}. During the talk of P.~Drell~\cite{drell} we have
seen many exciting new experimental results on $B$-decays, and the
flow of new data will continue for many years from existing and new
facilities. The main theoretical difficulty in interpreting the
experimental data is the control of the non-perturbative strong
interaction effects, and this problem is the main focus of my lecture.

For some of the physical quantities discussed below I will summarize
status of lattice results; these results have been taken from the
recent review written together with J.~Flynn~\cite{fs}, where more
details can be found. The lattice formulation of QCD, together with
large scale numerical simulations, enables one to calculate
non-perturbative QCD effects, from first principles, with no model
parameters or assumptions. The precision of the calculations is
limited, however, by the available computing resources, and much
effort is being devoted to reducing the systematic errors.  These
uncertainties, and the theoretical and computational efforts to
control and reduce them, have been discussed at this conference in
some detail by M.~Luscher~\cite{luscher}. I will therefore limit my
comments to the specific computations I am discussing.

The plan for the talk is as follows. In the next section I will review
the status of lattice calculations of the leptonic decay constants of
heavy mesons and of the amplitudes for the important process of
$B$-$\bar B$ mixing. Sec.~\ref{sec:semilexl} contains a review of
exclusive semileptonic decays of $B$-mesons which are being used to
extract the $V_{ub}$ and $V_{cb}$ CKM matrix elements. Inclusive
semileptonic decays are discussed very briefly in
sec.~\ref{sec:incsemi}, where the recent suggestions to determine
$V_{ub}$ by measuring the hadronic invariant mass spectrum are
outlined. I then digress from the mainstream of the presentation, to
discuss the problem of power corrections to hard scattering and decay
processes in general, and in the heavy quark effective theory (hqet) in
particular (sec.~\ref{sec:power}).  The next two sections contain
discussions of inclusive (sec.~\ref{sec:nlinclusive}) and exclusive
(sec.~\ref{sec:nlexc}) nonleptonic decays. In particular in
sec.~\ref{sec:nlinclusive} I will discuss two topics which have
received much attention lately, viz. the lifetimes of beauty hadrons
and the semileptonic branching ratio of $B$-mesons. Finally
sec.~\ref{sec:concs} contains some conclusions. $CP$-violation in
$B$-decays, which is perhaps the principal motivation for the coming
generation of $b$-factories, is the subject of the talk of
Y.~Nir~\cite{nir}, and will not be discussed below.

\section{Leptonic Decays and $B^0$--$\bar B^0$ Mixing}
\label{sec:fb}

I start with a discussion of the calculations of the leptonic decay
constants and of the $B$-parameter of $B$-$\bar B$ mixing.

\subsection{Leptonic Decay Constants}
\label{subsec:fb}
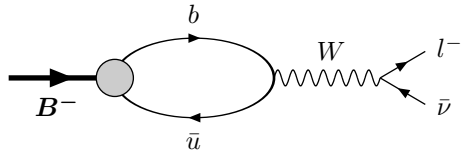
\begin{figure}[ht]
\begin{center}
\begin{picture}(180,50)(0,15)
\SetWidth{2}\ArrowLine(10,41)(50,41)
\SetWidth{0.5}
\Oval(80,41)(15,30)(0)
\ArrowLine(79,56)(81,56)\ArrowLine(81,26)(79,26)
\GCirc(50,41){7}{0.8}
\Photon(110,41)(150,41){3}{7}
\ArrowLine(150,41)(167,51)\ArrowLine(167,31)(150,41)
\Text(20,35)[tl]{\boldmath$B^-$}
\Text(80,62)[b]{$b$}\Text(80,20)[t]{$\bar u$}
\Text(172,53)[l]{$l^-$}\Text(172,30)[l]{$\bar\nu$}
\Text(132,48)[b]{$W$}
\end{picture}
\caption{Diagram representing the leptonic decay of the $B$-meson.}
\label{fig:fb}
\end{center}
\end{figure}

The decay constant of a meson is a single number which contains all
the information about the non-perturbative QCD effects in leptonic
decays of the meson (see e.g. Fig.~\ref{fig:fb}). Parity symmetry
implies that only the axial component of the $V-A$ weak current
contributes to the decay, and Lorentz invariance that the matrix
element of the axial current is proportional to the momentum of the
meson (with the constant of proportionality defined to be the decay
constant):
\begin{equation}
\langle 0|A_\mu(0)|B(p)\rangle\equiv i f_B p_\mu\ ,
\label{eq:fbdef}
\end{equation}
and similarly for the $D$-meson. Knowledge of $f_B$ would allow us not
only to predict the rates for leptonic decays, but would also be very
useful for describing $B$-$\bar B$ mixing as explained below,
as well as for our understanding of other processes in $B$-physics,
particularly those for which ``factorization'' turns out to be a
useful approximation.

Many lattice groups have evaluated, or are evaluating, $f_D$ and
$f_B$. Our view of the current status of the
calculations can be summarized by the following values for the decay
constants and their ratios~\cite{fs}: 
\begin{equation}
\begin{array}{ll}
\fd  =  200 \pm 30\mev\,,\ \ \ 
&\ \ \ \fds  =  220 \pm 30\mev\,,\label{eq:fdbest}\\  
\fb  =  170 \pm 35\mev\,,\ \ \ 
&\ \ \  \fbs =  195 \pm 35\mev\,,\label{eq:fbbest}\\ 
\fds/\fd  =  1.10 \pm 0.06\,,\ \ \    
&\ \ \ \fbs/\fb  =  1.14 \pm 0.08 \label{eq:fbs/fbbest}\ .  
\end{array}\end{equation}
All the results are presented using a normalization for which $f_{\pi^+}
\simeq 131\mev$.

In the absence of experimental results on the decay constants of the
$B$-meson, the results above represent our best estimates of this
quantity. During the past three years or so, experimental measurements
of $f_{D_s}$ have been made by several groups, and at this years {\it
Heavy Flavours} conference~\cite{nippe}, the raporteur summarized the
results by:
\begin{equation}
f_{D_s} = \left( 251\pm 30\right)\mev\ .
\label{eq:fdsexp}\end{equation}
The agreement of the lattice {\it predictions} with the experimental
result is very satisfying, and gives us confidence in the results
for the other decay constants.

The results for the decay constants have generally remained stable for
many years, but the quoted errors have not decreased significantly in
this time. This is because the errors are dominated by systematic
effects, and it is difficult to decrease the uncertainties due to the
quenched approximation (i.e. the neglect of vacuum polarization
effects) without performing reliable calculations with dynamical
fermions, which will still take several more years. For example, the
value of the lattice spacing typically varies by 10\% or so depending
on which physical quantity is used to calibrate the lattice
simulation. This variation is largely due to the use of the quenched
approximation. We therefore consider that $\pm$10\% is an irreducible
minimum error in decay constants computed in quenched simulations (and
a correspondingly larger error when physical quantities of higher
dimension are calculated).  The remainder of each error in
eqs.~(\ref{eq:fdbest}) and in the lattice results quoted below is
based on our estimates of the other uncertainties~\cite{fs},
particularly those due to discretization errors and the normalization
of the lattice composite operators.  Much successful work continues to
be done to reduce these uncertainties~\cite{luscher}, and the results
have remained stable during these improvements, adding significantly
to our confidence, but not reducing the quoted errors by very much.

\begin{figure}
\hbox to\hsize{\hss
\epsfxsize0.5\hsize\epsffile{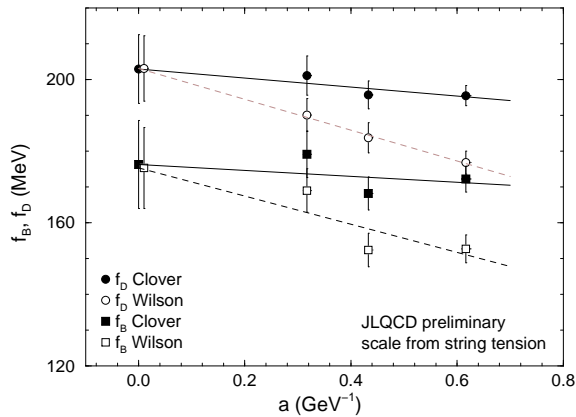}
\hss}
\caption[]{Continuum extrapolation of JLQCD~\cite{jlqcd:fb-lat97}
results (preliminary) for $f_D$ and $f_B$. The graph shows agreement
between results from two different lattice formulations when
extrapolated to zero lattice spacing. Open symbols denote an
unimproved Wilson action ($\csw=0$), filled symbols denote an improved
Sheikholeslami-Wohlert or Clover ($\csw=1$) action.}
\label{fig:jlqcd-fdfb}
\end{figure}

As an example of the efforts of some recent work aimed at reducing the
systematic uncertainties in the evaluation of the decay constants
consider Figure~\ref{fig:jlqcd-fdfb} which contains preliminary results
from the JLQCD collaboration~\cite{jlqcd:fb-lat97}.  The figure shows
results obtained at several different values of the lattice spacing $a$
and using two lattice formulations of the heavy quark action, the
standard Wilson formulation (but with a modified normalization of the
fields, using the so-called KLM-normalization~\cite{klm}) and with the
SW~\cite{sw} or ``clover'' improved action. The lattice was calibrated
using the string tension to determine the spacing. The feasibility of
obtaining (statistically) accurate results for a series of lattice
spacings and actions is a new development, largely made possible by the
increase in computing power. The figure also shows extrapolations to the
continuum ($a=0$) limit.  In this case the continuum values of $f_B$ and
$f_D$ obtained using the two formulations agree remarkably well; the
agreement is still acceptable (although not so remarkable) when
quantities other than the string tension are used to determine the
lattice spacing. We also note that in this case the dependence on the
lattice spacing is milder for the improved action as expected (although
further studies are needed to confirm whether this is a general
feature).

The MILC collaboration has began to study the effects of quenching,
and their very early findings are that the decay constants may be a
little larger (perhaps by about 10\%) when the effects of quark loops
are included~\cite{milc:fb-lat97}. Further work will be done in the
next few years to study this further.

\subsection{$B_B$-Parameter}
\label{subsec:bb}

In order to obtain information about the unitarity triangle from
experimental studies of $B$-$\bar B$ mixing, one needs to control
the non-perturbative QCD effects which are contained in the matrix
element of the $\Delta B{=}2$ operator:
\begin{equation}
M(\mu) = \langle\bar B^0 |\, \bar b \gamma_\mu (1{-}\gamma_5) q\,
  \bar b \gamma^\mu (1{-}\gamma_5) q \,| B^0 \rangle, 
\label{eq:mbbar}
\end{equation}
where the light quark $q {=} d$ or $s$ 
The argument $\mu$ implies that the operator has been
renormalized at the scale $\mu$. It is conventional to introduce the
$\bb$-parameter through the definition
\begin{equation}
M(\mu) = \frac{8}{3}\, f_B^2 m_B^2 \bb(\mu).
\label{eq:bbdef}
\end{equation}
The dimensionless quantity $\bb$ is better-determined than $f_B$ in
lattice calculations, so that the theoretical uncertainties in the
value of the matrix element $M$, needed for phenomenology,
are dominated by our ignorance of $f_B$.

$\bb(\mu)$ is a scale dependent quantity for which lattice results are
most often quoted after translation to the $\overline\mathrm{MS}$
scheme. The next-to-leading order (NLO) renormalization group
invariant $B$--parameter ($\rgbb^{\mathrm{nlo}}$) is defined by
\begin{equation}
\rgbb^{\mathrm{nlo}} = \alpha_s(\mu)^{-2/\beta_0}
 \left( 1 + {\alpha_s(\mu)\over4\pi} J_{n_f} \right) \bb(\mu),
\label{eq:bbhat}
\end{equation}
where $\beta_0 = 11 - 2n_f/3$ and $J_{n_f}$ is obtained from the
one- and two-loop anomalous dimensions of the $\Delta B{=}2$ operator
by~\cite{bjw:bb},
\begin{equation}
J_{n_f} = {1\over2\beta_0} \left( \gamma_0 {\beta_1\over\beta_0} -
 \gamma_1 \right),
\end{equation}
with $\beta_1 = 102 - 38 n_f/3$, $\gamma_0 = 4$ and $\gamma_1 = -7 + 4
n_f/9$. In the discussion below we choose $\mu=m_b$.

A number of groups have evaluated $\hat B$ in quenched lattice
simulations~\cite{bb}, from which we deduce the preferred
value~\cite{fs}
\begin{equation}
\rgbb^{\mathrm{nlo}} = 1.4(1)\ .
\label{eq:bbbest}
\end{equation}
The relevant quantity for $B$--$\bar B$ mixing is $f_B^2 \rgbb$. Taking
the result in eq.~(\ref{eq:bbbest}) above for $\rgbb^{\mathrm{nlo}}$
with $f_B = 170(35)\mev$ from eq.~(\ref{eq:fbbest}) gives
\begin{equation}
f_B\sqrt{\rlap{$\phantom{\hat B}$}\smash{\rgbb^{\mathrm{nlo}}}} =
201(42)\mev \label{eq:xibbest}
\end{equation}
as our lattice estimate. 
For the phenomenologically important quantity $\xi$, which relates the
amplitudes for $B_d-\bar B_d$ and $B_s-\bar B_s$ mixing, we
find~\cite{fs}:
\begin{equation}
\xi\equiv {\fbs \sqrt{\rlap{$\phantom{\hat B}$}\smash{\rgbbs}}  \over
 \fbd \sqrt{\rlap{$\phantom{\hat B}$}\smash{\rgbbd}}}
= 1.14(8)\ .
\label{eq:xibest}\end{equation}

\section{Exclusive Semileptonic Decays of $B$-Mesons}
\label{sec:semilexl}

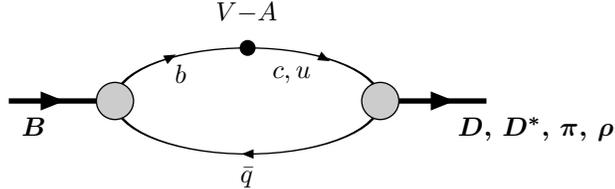
\begin{figure}[ht]
\begin{center}
\begin{picture}(180,70)(20,10)
\SetWidth{2}\ArrowLine(10,41)(43,41)
\Text(15,35)[tl]{\boldmath$B$}
\SetWidth{0.5}
\Oval(100,41)(20,50)(0)
\SetWidth{2}\ArrowLine(157,41)(190,41)
\Text(180,35)[tl]{\boldmath$D,\,D^*,\,\pi,\,\rho$}
\SetWidth{0.5}
\Vertex(100,61){3}
\GCirc(50,41){7}{0.8}\GCirc(150,41){7}{0.8}
\Text(75,48)[b]{$b$}\Text(117,48)[b]{$c,u$}
\Text(100,16)[t]{$\bar q$}
\Text(100,71)[b]{$V{-}A$}
\ArrowLine(101,21)(99,21)
\ArrowLine(70,57)(72,58)\ArrowLine(128,57.5)(130,56.5)
\end{picture}
\caption{Diagram representing the semileptonic decay of the $B$-meson.
$\bar q$ represents the light valence antiquark, and the black circle
represents the insertion of the $V$--$A$ current with the appropriate
flavour quantum numbers.}
\label{fig:sl}
\end{center}
\end{figure}

Semileptonic decays of $B$-mesons 
are currently being used to determine the $V_{cb}$ and $V_{ub}$
CKM-matrix elements. Exclusive decays are represented by the diagram in
Figure~\ref{fig:sl}. It is convenient to use space-time symmetries to
express the matrix elements in terms of invariant form factors (using
the helicity basis for these as defined below).  When the final state
is a pseudoscalar meson $P$, parity implies that only the vector
component of the $V-A$ weak current contributes to the decay, and
there are two independent form factors, $f^+$ and $f^0$, defined by
\begin{equation}
\langle P(k)| V^\mu|B(p)\rangle  = 
  f^+(q^2)\left[(p+k)^\mu -
  \frac{m_B^2 - m_P^2}{q^2}\,q^\mu\right] 
\mbox{} + f^0(q^2)\,\frac{m_B^2 - m_P^2}{q^2}\,q^\mu\ ,
\label{eq:ffpdef}
\end{equation}
where $q$ is the momentum transfer, $q=p{-}k$. 
When the final-state hadron is a vector
meson $V$, there are four independent form factors:
\begin{eqnarray}
\langle V(k,\varepsilon)| V^\mu|B(p)\rangle & = &
  \frac{2V(q^2)}{m_B+m_V}\epsilon^{\mu\gamma\delta\beta}
  \varepsilon^*_\beta p_\gamma k_\delta \label{eq:ffvvdef}\\ 
\langle V(k,\varepsilon)| A^\mu|B(p)\rangle  & = &  
  i (m_B {+} m_V) A_1(q^2) \varepsilon^{*\,\mu}
 - 
  i\frac{A_2(q^2)}{m_B{+}m_V} \varepsilon^*\!\dotprod p\,
  (p {+} k)^\mu 
\mbox{} + i \frac{A(q^2)}{q^2} 2 m_V 
  \varepsilon^*\!\dotprod p\, q^\mu , \label{eq:ffvadef}
\end{eqnarray}
where $\varepsilon$ is the polarization vector of the final-state
meson, and $q = p{-}k$.  Below we shall also discuss the form factor
$A_0$, which is given in terms of those defined above by $A_0 = A +
A_3$, with
\begin{equation}
A_3 = \frac{m_B + m_V}{2 m_V}A_1 - 
\frac{m_B - m_V}{2 m_V}A_2\ .
\label{eq:a3def}
\end{equation} 

\subsection{$B\to D^*$ and $B\to D$ Decays}
\label{subsec:vcbexl}

Exclusive $B\to D^*$ and, more recently, $B \to D$ semileptonic decays 
are used to determine
the $V_{cb}$ element of the CKM matrix. Theoretically they are 
relatively simple to consider, since heavy quark symmetry
implies that the six form factors are related, and that there is only
one independent form factor $\xi(\omega)$, specifically:
\begin{equation}
f^+(q^2)  =  V(q^2) = A_0(q^2) = A_2(q^2) 
 =  \left[1 - 
\frac{q^2}{(M_B + M_D)^2}\right]^{-1} A_1(q^2)\equiv\frac{M_B+M_D}
{2\sqrt{M_BM_D}}\,\xi(\omega)\ ,
\label{eq:iw}\end{equation}
where $\omega = v_B\cdot v_D$. Here the label $D$ represents the $D$-
or $D^*$-meson as appropriate. In this leading approximation the
pseudoscalar and vector mesons are degenerate. The unique form factor
$\xi(\omega)$, which contains all the non-perturbative QCD effects, is
called the Isgur--Wise (IW) function. Vector current conservation
implies that the IW-function is normalized at zero recoil, i.e. that
$\xi(1) =1$. This property is particularly important in the extraction
of the $V_{cb}$ matrix element.

The relations in eq.~(\ref{eq:iw}) are valid up to perturbative and
power corrections. The precision with which $V_{cb}$ can be extracted
is limited by the theoretical uncertainties in the evaluation of these
corrections. Nevertheless we are in the fortunate situation that it is
uncertainties in {\it corrections} (which are therefore small) which
control the precision.

The decay distribution for $B\to D^*$ decays can be written as:
\begin{equation}
\frac{d\Gamma}{d\omega} =  \frac{G_F^2}{48\pi^3}
(M_B-M_{D^*})^2 M_{D^*}^3 \sqrt{\omega^2 -1}\,(\omega+1)^2\,
\left[ 1 + \frac{4\omega}{\omega + 1} 
\frac{M_B^2 - 2\omega M_BM_{D^*}+M_{D^*}^2}{(M_B-M_{D^*})^2}\right]
|V_{cb}|^2\, {\cal F}^2(\omega)\ ,
\label{eq:distr}\end{equation}
where ${\cal F}(\omega)$ is the IW-function combined with perturbative
and power corrections. It is convenient theoretically to
consider this distribution near $\omega = 1$. In this case $\xi(1) =
1$, and there are no $O(1/m_Q)$ corrections (where $Q= b$ or $c$) by
virtue of Luke's theorem~\cite{luke}, so that the expansion of ${\cal
  F}(1)$ begins like:
\begin{equation}
{\cal F}(1) = \eta_A\left( 1 + 0\,\frac{\Lambda_{QCD}}{m_Q} + 
c_2\frac{\Lambda^2_{QCD}}{m_Q^2} + \cdots\right)\, ,
\label{eq:fexpansion}\end{equation}
where $\eta_A$ represents the perturbative corrections.
The one-loop contribution to $\eta_A$ has been known for some time now,
whereas the two-loop contribution was evaluated last year, with the
result~\cite{czarnecki}:
\begin{equation}
\eta_A = 0.960\pm 0.007\ ,
\end{equation}
where we have taken the value of the two loop contribution as an
estimate of the error.

The power corrections are much more difficult to estimate reliably.
Neubert has recently combined the results of
refs.~\cite{fn,mannel,suv} to estimate that the $O(1/m_Q^2)$ terms in
the parentheses in eq.~(\ref{eq:fexpansion}) are about $-0.055\pm
0.025$ and that ${\cal F}(1) = 0.91 (3)\,$~\cite{neubertcernschool}.
Bigi, Shifman and Uraltsev~\cite{bsu}, consider the uncertainties to
be bigger and obtain $0.91(6)$. Combining the latter, more cautious,
theoretical value of ${\cal F}(1)$, with the experimental
result~\cite{diciaccio} ${\cal F}(1)|V_{cb}|= (34.3\pm 1.6)10^{-3}$,
readily gives $|V_{cb}|=(37.7\pm 1.8_{exp}\pm 2.5_{th})10^{-3}$.
In considering this result, the fundamental question that
has to be asked is whether the power corrections are sufficiently
under control. This will be discussed in more detail in
section~\ref{sec:power}.  

Having discussed the theoretical status of the normalization ${\cal
  F}(1)$, let us now consider the shape of the function ${\cal
  F}(\omega)$, near $\omega = 1$. A theoretical understanding of the
shape would be useful to guide the extrapolation of the experimental
data, and also as a test of our understanding of the QCD effects. We
expand ${\cal F}$ as a power series in $\omega -1$:
\begin{equation}
{\cal F}(\omega) = {\cal F}(1)\, \left[1 - \hat\rho^2\,(\omega -1)
+\hat c\, (\omega -1 )^2 + \cdots\right]\ ,
\label{eq:ftaylor}\end{equation}
where~\cite{neubertcernschool}
\begin{equation}
\hat\rho^2 = \rho^2 + (0.16\pm 0.02) + \mbox{power corrections}\ ,
\label{eq:rhohat}\end{equation}
and $\rho^2$ is the slope of the IW-function. 

Recently, using analyticity and unitarity properties of the amplitudes,
as well as the heavy quark symmetry, Caprini and Neubert have derived an
intriguing result for the curvature parameter $\hat c\,$~\cite{caprini},
$\hat c \simeq 0.66\, \hat\rho^2 - 0.11$. This result would have
effectively removed one of the parameters from the extrapolation of the
experimental data to $\omega=1$. The derivation of this relation has
been criticized in ref.~\cite{lebed}, primarily for discarding
sub-threshold ($B_c$) contributions, and the expectation is that
the corrected relations give somewhat weaker information. Caprini et al.,
are preparing a revised slope-curvature relation~\cite{caprini2};
preliminary indications were presented in ref.~\cite{mnjerusalem}.

Finally in this section I consider $B\to D$ semileptonic decays, which
are beginning to be measured experimentally with good precision (see
refs.~\cite{artuso,lkg:ichep96} and references therein). The decay
distribution can be written as:
\begin{equation}
\frac{d\Gamma}{d\omega}=\frac{G_F^2}{48\pi^3}(M_B+M_D)^2
M_D^3(\omega^2-1)^{\frac{3}{2}}|V_{cb}|^2{\cal G}^2(\omega)\ ,
\label{eq:distr3}\end{equation}
where again ${\cal G}(\omega)$ is the IW-function combined with
perturbative and power corrections.  Theoretically the first
complication is that the $1/m_Q$ corrections to ${\cal G}(1)$ do not
vanish. However, as pointed out by Shifman and Voloshin~\cite{sv},
these corrections would vanish in the limit in which the $b$ and
$c$-quarks are degenerate, and hence are suppressed.  Ligeti, Nir, and
Neubert estimate the $1/m_Q$ corrections to be between approximately
$-$1.5\% to +7.5\%~\cite{lnn}. The $1/m_Q^2$ corrections for this decay
have not yet been studied systematically. A recent theoretical estimate
for ${\cal G}(1)$ gives $0.98\pm0.07$~\cite{mnjerusalem}.

\subsection{$B\to \rho$ and $B\to \pi$ Decays}
\label{subsec:vubexc}

In this subsection we consider the heavy-to-light semileptonic decays
$B\to\rho$ and $B\to\pi$ which are now being used experimentally to
determine the $V_{ub}$ matrix element~\cite{lkg:ichep96,jrp:ichep96}.
Heavy quark symmetry is less predictive for heavy-to-light decays than
for heavy-to-heavy ones.  In particular, there is no normalization
condition at zero recoil corresponding to the relation $\xi(1)=1$,
which is so useful in the extraction of $V_{cb}$. The lack of such a
condition puts a premium on the results from nonperturbative
calculational techniques, such as lattice QCD or light-cone sum rules.
Heavy quark symmetry does, however, give useful scaling laws for the
behaviour of the form factors with the mass of the heavy quark ($m_Q$)
at fixed $\omega$:
\begin{equation}
f^+, A_0,A_2,V\sim\sqrt{m_Q}\ ;\ A_1,f^0\sim\frac{1}{\sqrt{m_Q}}\ ;\ 
A_3\sim m_Q^{\frac{3}{2}}\ .
\label{eq:hqscaling}\end{equation}
These scaling relations are particularly useful in lattice simulations,
where the masses of the quarks are varied.  Moreover, the heavy quark
spin symmetry relates the $B\to V$ matrix elements~\cite{iw:hqet,gmm}
(where $V$ is a light vector particle) of the weak current and magnetic
moment operators, thereby relating the amplitudes for the two physical
processes $\btorho$ and $\btokstargamma$, up to $SU(3)$ flavour symmetry
breaking effects.  These relations also provide important checks on
theoretical, and in particular on lattice, calculations.

Recent work includes detailed lattice studies by several groups,
but in particular by the UKQCD collaboration, who try to exploit 
all possible symmetries in order to optimize the available 
information about the form factors and calculations using light-cone
sum rules.

\subsection{Lattice Calculations of $B\to\rho,\pi$ Semileptonic Decays
and of the Rare Decay $\btokstargamma$}
\label{subsec:latbtorho}

The techniques required to extract the form-factors for $\btopi$ and
$\btorho$ semileptonic decays are very similar to those used to compute
the short distance contribution to the rare radiative decay
$\btokstargamma$, so I consider them together. For completeness, I
define here form factors for the matrix element of the magnetic moment
operator responsible for this decay:
\begin{equation}
\langle K^*(k,\varepsilon) | \bar{s} \sigma_{\mu\nu} q^\nu b_R
 | B(p) \rangle
  =  \sum_{i=1}^3 C^i_\mu T_i(q^2),
\label{eq:bkstarME}
\end{equation}
where $q=p{-}k$, $\varepsilon$ is the polarization vector of the $K^*$ and
\begin{eqnarray}
C^{1}_\mu & = &
  2 \epsilon_{\mu\nu\lambda\rho} \varepsilon^{*\,\nu} p^\lambda k^\rho, \\
C^{2}_\mu & = &
  \varepsilon^*_\mu(m_B^2 - m_{K^*}^2) - \varepsilon\dotprod q (p+k)_\mu, \\
C^{3}_\mu & = & \varepsilon^*\!\dotprod q
  \left( q_\mu - \frac{q^2}{m_B^2-m_{K^*}^2} (p+k)_\mu \right).
\end{eqnarray}
$T_3$ does not contribute to the physical $\btokstargamma$ amplitude
for which $q^2=0$, and $T_1(0)$ and $T_2(0)$ are related by,
\begin{equation}
T_1(q^2{=}0) = i T_2(q^2{=}0).
\end{equation}
Hence, for the process $\btokstargamma$, we need to determine $T_1$
and/or $T_2$ at the on-shell point $q^2{=}0$.

\begin{figure}
\unit 0.6\hsize
\hbox to\hsize{\hss\vbox{\offinterlineskip
\epsfxsize\unit
\epsffile[-25 54 288 232]{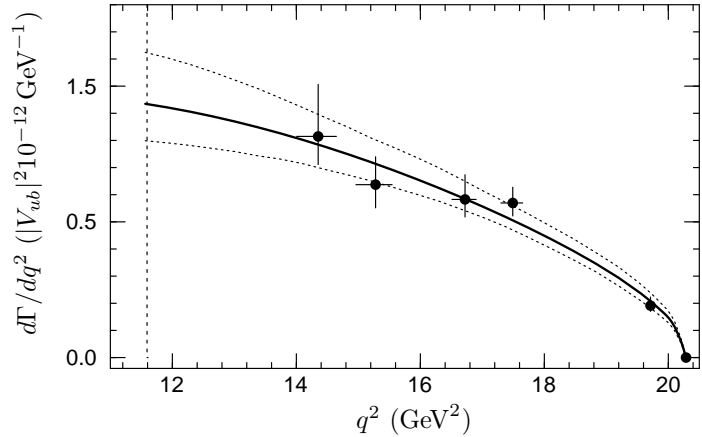}
\point 0 0.56 {\begin{sideways}
$d\Gamma/dq^2\ (|V_{ub}|^2 10^{-12}\gev^{-1})$\end{sideways}}
}\hss}
\hbox to\hsize{\hss
\hbox to\unit{\kern0.5\unit$q^2\ (\!\gev^2)$\hfill}\hss}
\caption[]{Differential decay rate as a function of $q^2$ for the
semileptonic decay $\bar B^0\to\rho^+l^-\bar\nu_l$, taken
from~\cite{ukqcd:btorho}. Points are measured lattice data, solid
curve is fit from eq.~(\ref{eq:distr2}) with parameters given in
eq.~(\ref{eq:ab-vals}). The dashed curves show the variation from the
statistical errors in the fit parameters. The vertical dotted line
marks the charm endpoint.}
\label{fig:vub}
\end{figure}

From lattice simulations we can only obtain the form factors for part of
the physical phase space for all the decays.  In order to control
discretization errors we require that the three-momenta of the $B$,
$\pi$ and $\rho$ mesons be small in lattice units. This implies that we
determine the form factors at large values of momentum transfer $q^2 =
(p_B-p_{\pi,\rho,K^*})^2$. Experiments can already reconstruct exclusive
semileptonic $b\to u$ decays (see, for example, the review
in~\cite{jrp:ichep96}) and as techniques improve and new facilities
begin operation, we can expect to be able to compare the lattice form
factor calculations directly with experimental data at large $q^2$. A
proposal in this direction was made by UKQCD~\cite{ukqcd:btorho} for
$\btorho$ decays. To get some idea of the precision that might be
reached, they parametrize the differential decay rate distribution near
$\qsqmax$ by:
\begin{equation}
\frac{d\Gamma(\btorho)}{dq^2}
 =  10^{-12}\,\frac{G_F^2|V_{ub}|^2}{192\pi^3M_B^3}\,
q^2 \, \lambda^{\frac{1}{2}}(q^2)
 \, a^2\left( 1 + b(q^2{-}\qsqmax)\right),
\label{eq:distr2}
\end{equation}
where $a$ and $b$ are parameters, and the phase-space factor $\lambda$
is given by $\lambda(q^2) = (m_B^2+m_\rho^2 - q^2)^2 - 4
m_B^2m_\rho^2$. The constant $a$ plays the role of the IW function
evaluated at $\w=1$ for heavy-to-heavy transitions, but in this case
there is no symmetry to determine its value at leading order in the
heavy quark effective theory. UKQCD obtain~\cite{ukqcd:btorho}
\begin{equation}
\begin{array}{ccc}
a = 4.6 \err{0.4}{0.3} \pm 0.6 \gev & \ \mbox{and}\ &
b = (-8 \err 46) \times 10^{-2} \gev^2.
\end{array} \label{eq:ab-vals} \end{equation}
The fits are less sensitive to $b$, so it is less well-determined. The
result for $a$ incorporates a systematic error dominated by the
uncertainty ascribed to discretization errors and would lead to an
extraction of $|V_{ub}|$ with less than 10\% statistical error and
about 12\% systematic error from the theoretical input.  The
prediction for the $d\Gamma/dq^2$ distribution based on these numbers
is presented in Fig.~\ref{fig:vub}.  With sufficient experimental data an
accurate lattice result at a single value of $q^2$ would be sufficient
to fix $|V_{ub}|$.

In principle, a similar analysis could be applied to the decay
$\btopi$. However, UKQCD find that the difficulty  of performing the
chiral extrapolation to a realistically light pion from the unphysical
pions used in the simulations makes the results less certain. 
The $B\to\pi$ decay also has a smaller fraction of events at high
$q^2$, so it will be more difficult experimentally to extract
sufficient data in this region for a detailed comparison.

\begin{figure}
\unit=0.7\hsize
\hbox to\hsize{\hss
\vbox{\offinterlineskip
\epsfxsize=\unit\epsffile[-20 50 288 236]{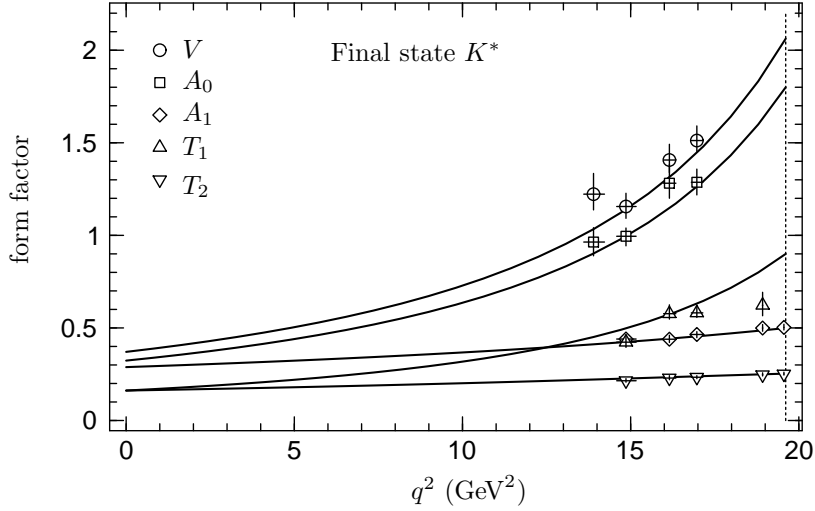}
\point 0.4 0.54 {{\rm Final\ state\ }K^*}
\point 0.215 0.543 V
\point 0.215 0.505 {A_0}
\point 0.215 0.465 {A_1}
\point 0.215 0.422 {T_1}
\point 0.215 0.374 {T_2}
\point 0 0.425 {\begin{sideways}{form factor}\end{sideways}}
\hbox to\unit{\kern0.5\unit$q^2\ (\!\gev^2)$\hfill}
}\hss}
\caption[]{UKQCD~\cite{ukqcd:hlfits} fit to the lattice predictions
for $A_0$, $A_1$, $V$, $T_1$ and $T_2$ for a $K^*$ meson final state
assuming a pole form for $A_1$. $A_2$ is not reliably extracted from
the lattice data so is not used in the fit. The dashed vertical line
indicates $\qsqmax$.
\label{fig:ukqcd-kstarfit}}
\end{figure}

We would also like to know the full $q^2$ dependence of the form
factors, which involves a large extrapolation in $q^2$ from the high
values where lattice calculations produce results, down to $q^2=0$. In
particular the radiative decay $\btokstargamma$ occurs at $q^2=0$, so
that existing lattice simulations cannot make a direct calculation of
the necessary form factors. In order to determine the form factors at
lower values of $q^2$ from their measurements the lattice
collaborations, and the UKQCD collaboration in particular, exploit a
number of important constraints. I now briefly outline these in turn:
\begin{itemize}
\item 
An interesting contribution to the problem of extrapolation to low $q^2$
has been suggested by Lellouch~\cite{lpl:bounds} for $\btopi$ decays.
Using dispersion relations constrained by UKQCD lattice results at large
values of $q^2$ and kinematical constraints at $q^2=0$, one can tighten
the bounds on form factors at all values of $q^2$. This technique relies
on perturbative QCD in evaluating one side of the dispersion relations,
together with general properties of field theory, such as unitarity,
analyticity and crossing. It provides model-independent results which
are illustrated in Fig.~\ref{fig:btopi-bounds}. The results
(at 50\% CL -- see Ref.~\cite{lpl:bounds} for details) are
\begin{equation}
\begin{array}{ccc}
f^+(0) = 0.10\hbox{--}0.57 & \mbox{and} &
\Gamma(\btopi) = 4.4\hbox{--}13\, |V_{ub}^2| \, \mathrm{ps}^{-1}.
\end{array}\end{equation}
Unfortunately these bounds are not very restrictive when constrained by
existing lattice data. In principle, this method can be applied to
$B\to\rho$ decays also, but is more complicated there, and the
calculations have yet to be performed. Recently,
Becirevic~\cite{becirevic:bksbounds} has applied the method for the
process $\btokstargamma$, using lattice results from the APE
collaboration as constraints. However, he has not applied the kinematic
constraint at $q^2=0$ and the resulting bounds are not informative: they
become so, however, once he uses a light-cone sum rule evaluation of the
form-factors for the process $\btokstargamma$ as an additional
constraint. These dispersive methods can be used with other approaches
in addition to lattice results and sum rules, such as quark models, or
even in direct comparisons with experimental data, to check for
compatibility with QCD and to extend the range of results.
\begin{figure}
\hbox to\hsize{\hss\vbox{\offinterlineskip
\unit=0.5\hsize
\epsfxsize=\unit\epsffile{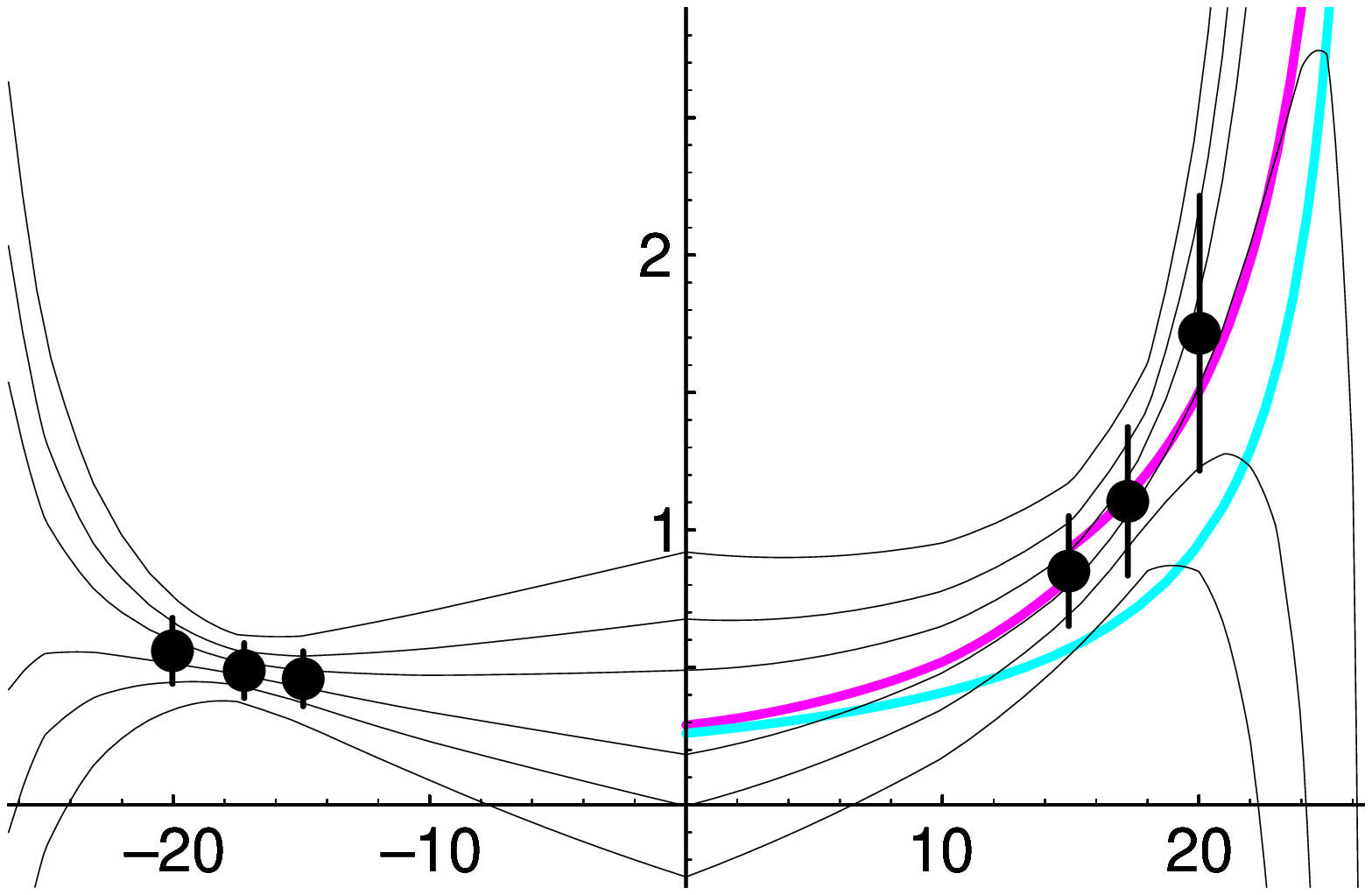}
\point 0.55 0.74 {f^+(q^2)}
\point 0.3 0.74 {f^0(|q^2|)}
\point 0.82 0.07 {q^2\ (\!\gev^2)}
}\hss}
\caption[]{Bounds on $f^+$ and $f^0$ for $\btopi$ from dispersive
constraints~\cite{lpl:bounds}. The data points are from
UKQCD~\cite{ukqcd:hlff}, with added systematic errors. The pairs of
fine curves are, outermost to innermost, 95\%, 70\% and 30\% bounds.
The upper and lower shaded curves are
light-cone~\protect\cite{bbkr:B-Bstar-pi-couplings} and
three-point~\protect\cite{PaB93} sum rule results respectively.}
\label{fig:btopi-bounds}
\end{figure}
\item UKQCD make use of the kinematic constraints on the form factors at
$q^2=0$:
\begin{equation}
f^+(0) = f^0(0), \qquad T_1(0) = iT_2(0), \qquad A_0(0) = A_3(0)\ .
\label{eq:ffkinconstraints}
\end{equation}
\item In spite of all the constraints, model input is required to
guide $q^2$ extrapolations.  We can ensure that any assumed
$q^2$-dependence of the form factors is consistent with the requirements
imposed by heavy quark symmetry, as shown in (\ref{eq:hqscaling}),
together with the kinematical relations of
eq.~(\ref{eq:ffkinconstraints}). UKQCD also verify that
the expected relations between form factors of the different
processes  (at fixed $\w$)
\begin{equation}
A_1\big(q^2(\w)\big)=2iT_2\big(q^2(\w)\big),\quad
 V\big(q^2(\w)\big)=2T_1\big(q^2(\w)\big),
\label{eq:a1t2vt1hqs}
\end{equation}
are indeed well satisfied in the infinite mass limit.
Even with all these constraints, however,
current lattice data do not by themselves distinguish a preferred
$q^2$-dependence. Fortunately, more guidance is available from
light-cone sum rule analyses~\cite{abs,patricia} which lead to scaling
laws for the form factors at fixed (low) $q^2$ rather than at fixed $\w$
as in eq.~(\ref{eq:hqscaling}). In particular all form factors scale like
$M^{-3/2}$ at $q^2=0$:
\begin{equation} f(0) \Theta = M^{-3/2} \gamma_f \left( 1
+ {\delta_f\over M} +
 {\epsilon_f\over M^2} + \cdots \right),
\label{eq:scalingm}\end{equation}
where $f$ labels the form factor, M is the mass of the heavy-light meson
and $\Theta$ is a calculable leading logarithmic correction.
\end{itemize}
A combined fit to the lattice data for the form factors
(obtained at large values of $q^2$) satisfying all these constraints
is shown in Fig.~\ref{fig:ukqcd-kstarfit}. The quark masses have been
chosen to correspond to the $K^*$ vector-meson. The figure
demonstrates the large extrapolation needed to reach $q^2=0$.

Our preferred results for $\btopi$ and $\btorho$ come from the UKQCD
constrained fits~\cite{ukqcd:hlfits}. Their best estimates of the total
rates are:
\begin{equation}
\begin{array}{ccc}
\Gamma(\bar B^0\to\pi^+e^-\bar\nu_e)  =  8.5\err{3.3}{1.4}\,
|V_{ub}|^2 ps^{-1} &\ \mbox{and}\ &
\Gamma(\bar B^0\to\rho^+e^-\bar\nu_e)  =  16.5\err{3.5}{2.3}\,
|V_{ub}|^2 ps^{-1}\ .\label{eq:btorhoresult}
\end{array}\end{equation}

There are also preliminary results for heavy-to-light form
factors from FNAL, JLQCD and a Hiroshima-KEK group (see the reviews
in~\cite{jmfstlouis,onogi:lat97}) and the different lattice
calculations are in agreement for the form factors at large
$q^2$ where they are measured.

For the form factor $T_1(0)$ of $\btokstargamma$ decays, the combined
fits give
\begin{equation}
T_1(0) = 0.16\errp21.
\label{eq:t0best}
\end{equation}
Using this value to evaluate the ratio (given at leading order in QCD
and up to $O(1/m_b^2)$ corrections~\cite{ciuchini94})
\begin{equation}
R_{K^*} = {\Gamma(\btokstargamma)\over\Gamma(b\to s\gamma)} =
  4\left(\frac{m_B}{m_b}\right)^3
  \left(1-\frac{m^2_{K^*}}{m^2_B}\right)^3 |T(0)|^2
\end{equation}
results in
\begin{equation}
R_{K^*} = 16\errp43 \%,
\end{equation}
which is consistent with the experimental result $18(7)\%$ from
CLEO~\cite{cleo:rkstar}. Discrepancies between $R_{K^*}$ calculated
using $T(0)$ and the experimental ratio
$\Gamma(\btokstargamma)/\Gamma(b\to s\gamma)$ could reveal the
existence of long-distance effects in the exclusive decay. It has been
proposed that these effects may be significant for the process
$\btokstargamma$~\cite{gp:longdist1}$^-$\cite{abs:longdist}, but
within the precision of the experimental and lattice results, there is
no evidence for them.

\subsection{Light Cone Sum Rule Studies of the Form Factors for
$\btopi$ and $\btorho$ decays}
\label{subsec:srbtorho}

The second technique which has been used to study heavy\,$\to$\,light
decays is QCD sum rules. Ball and Braun~\cite{babr} have recently
clarified the origin of the discrepancy in results obtained using
different types of sum rules for $B\to\rho$ semileptonic decays at
large recoil~\cite{PaB93,abs}. These authors explain why the standard
sum rules, in which one performs an Operator Product Expansion (OPE)
in terms of operators of increasing dimension, fail in this region of
phase space.  They stress the necessity of using light cone sum rules,
in which the contributions are classified by the components of
different ``twist" of the distribution amplitude of the $\rho$-meson
(strictly speaking of its moments).  This work exploits and extends
earlier studies~\cite{abs,cz,asy}. A careful investigation of the
dominant contributions to the form factors at large recoil yields the
scaling law in eq.~(\ref{eq:scalingm}) for the behaviour of the form
factors with the mass of the heavy quark at small fixed $q^2$ (rather
than fixed $\omega$).

\begin{figure}
\hbox to\hsize{\hss\vbox{\offinterlineskip
\unit=\hsize
\epsfxsize=\unit\epsffile{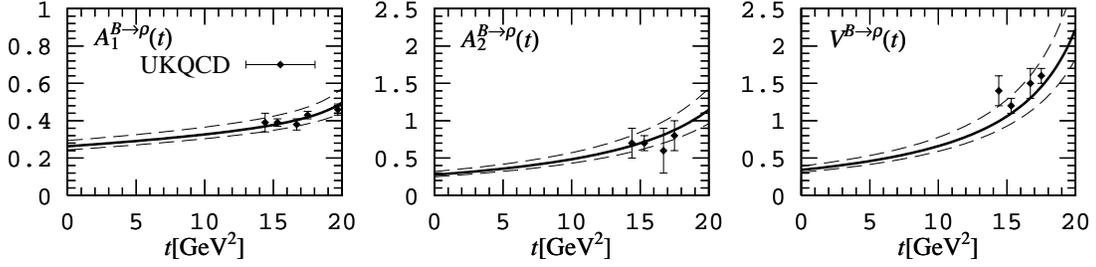}
}\hss}
\caption[]{Semileptonic $B$ form-factors as functions of $t=q^2$ obtained
from light-cone sum rules~\cite{babr}. The dashed lines give error
estimates and the points are from the lattice simulation of the 
UKQCD collaboration~\cite{ukqcd:btorho}.}
\label{fig:bb}
\end{figure}

The form factors derived in ref.~\cite{babr} are shown in
Fig.~\ref{fig:bb} together with the results from the UKQCD
collaboration at large values of $q^2$. In view of the uncertainties
in both sets of calculations, the agreement between the sum rule and
lattice results is remarkable. For the total rate, Ball and Braun find
\begin{equation}
\Gamma(\bar B^0\to\rho^+e^-\bar\nu_e) = (13.5\pm 1.0\pm 1.3\pm 0.6
\pm 3.6)
|V_{ub}|^2 ps^{-1}\ ,\label{eq:btorholcsr}
\end{equation}
to be compared to the lattice result in eq.(\ref{eq:btorhoresult}).
The second error in eq.~(\ref{eq:btorholcsr}) is due to the uncertainty
in the mass of the $b$-quark, the remaining errors are estimates of
various uncertainties in the light-cone sum rule calculation and in
the distribution amplitude of the $\rho$-meson.

\section{Inclusive Semileptonic Decays - $V_{ub}$}
\label{sec:incsemi}

The energy of the electron ($E_e$) in semileptonic $B$-decays is
limited kinematically to a very narrow window: 
\begin{equation}
\frac{m_B^2-m_D^2}{2m_B} \simeq 2.3\ \mathrm{GeV}\leq E_e
\leq\frac{m_B}{2}\simeq 2.6\ \mathrm{GeV}\ .
\label{eq:eewindow}\end{equation}
This window contains only a small fraction (about 10\%) of all the
$b\to u$ decays, and it is difficult to make theoretical predictions
for the spectrum.

Two groups have recently proposed to use the hadronic invariant mass
($M_h$) spectrum instead of the electron energy
spectrum~\cite{flw,dikeman}~\footnote{The use of the inclusive
hadronic energy spectrum for the determination of $V_{ub}$ had been
proposed in ref.~\cite{rey}.}. About 90\% of $b\to u$ decays satisfy
$M_h<M_D$.
The spectrum takes the form
\begin{equation}
\begin{array}{ccc}
\displaystyle
\frac{d\Gamma}{dM_X^2} = \frac{G_F^2m_b^5}{192\pi^3}|V_{ub}|^2S(y)
&\ \mbox{where}\ &
\displaystyle
y=\frac{M_X^2}{\bar\Lambda m_b}\ \ \mathrm{and}\  \
\bar\Lambda=m_B-m_b\ .
\label{eq:ylambdabar}\end{array}\end{equation}
The non-perturbative QCD effects are contained in $S(y)$, and specifically
in the parameters $a_i$ which can be expressed as matrix elements of 
composite operators:
\begin{equation}
\langle H(v)|\bar h_v (iD_{\mu_1}) (iD_{\mu_2})\cdots (iD_{\mu_n})
h_v| H(v)\rangle
= 2 m_Ba_n v_{\mu_1}v_{\mu_2} \cdots v_{\mu_n}\ ,
\label{eq:andef}\end{equation} 
where $a_0=1$, $a_1=0$ and $a_2$ is given in terms of the kinetic
energy of the heavy quark in the meson. 

It is proposed to determine $V_{ub}$ from the integral of the spectrum
up to some maximum hadronic mass; the precision will depend on the
value of the cut-off which can be attained experimentally (the
precision on $V_{ub}$ is estimated to be 10-20\% for values of the
cut-off from $M_D$ down to 1.5~GeV or so). Much theoretical and
experimental work is needed to extract the optimal results from this
method.

\section{Power Corrections}
\label{sec:power}

I now digress from the main discussion of $B$-decays to consider the
evaluation of power corrections to hard scattering and decay processes. 
Since there are many confusing statements in the literature, and because
the evaluation of higher order terms in the heavy quark expansion is
very important in $B$-physics, it may be useful to clarify some of the
key points.  Although the discussion is presented in the context of
$B$-decays, it can readily be applied to other processes for which an
OPE is useful and where the power corrections are given by the matrix
elements of higher-twist or higher-dimension operators. The approach
presented here was developed together with G.~Martinelli~\cite{ht},
where references to the key papers can be found.

Consider some physical quantity ${\cal P}$ for which the OPE allows us
to write the theoretical prediction in terms of an expansion in inverse
powers of $m_b$:
\begin{equation}
{\cal P} = C_0(m_b,\mu)\langle f|O_0(\mu)|i\rangle +
\frac{1}{m_b^n} C_1(m_b,\mu)\langle f|O_1(\mu)|i\rangle +\cdots
\label{eq:ope}\end{equation}
where the ellipsis represents higher order terms in the OPE which we
will not consider further.  For simplicity I assume here that there is
only a single operator in each of the first two orders of the
expansion; if this is not the case then there will be an additional
mixing of operators which only requires a minor modification of the
discussion below. For clarity of notation I suppress the dependence of
the Wilson coefficient functions ($C_i$) on the strong coupling
constant.

I assume that we are interested in evaluating the $O(1/m_b^n)$
corrections in ${\cal P}$. The theory of power corrections is very
delicate, since they are exponentially small compared with the terms of
the perturbation series for $C_0$\, ($1/m_b\sim\exp(-c/\alpha_s(m_b))$,
where $c$ is a constant). Thus in order to evaluate the power
corrections, we need to control the exponentially small tail of the
perturbation series for $C_0$. This raises the problems of the Borel
summability and uniqueness of the sum of this perturbation series, and
the question of renormalon ambiguities which I will not discuss further
here~\cite{david,mueller}. 

Frequently, renormalization schemes based on the dimensional
regularization of ultraviolet divergences are used to define
renormalized operators (e.g. the $\MSbar$ scheme). One consequence of
the problems mentioned above is that higher dimensional operators (such
as $O_1$) are not uniquely defined in the $\MSbar$ scheme; the remaining
ambiguity in the matrix elements of $O_1$ is of $O(\Lambda_{QCD}^n)$,
i.e. of the same order as the matrix elements themselves). The
exceptions to this are operators whose matrix elements give the leading
contribution to some physical quantity, e.g. the dimension 5 chromomagnetic
operator
\begin{equation}
\begin{array}{ccc}
\displaystyle
\lambda_2  =  \frac{1}{3}\frac{\langle B|\bar
h\frac{1}{2}\sigma_{ij}G^{ij}h|B\rangle}{2M_B} &
\mbox{where}
&
\displaystyle
4\lambda_2  =  M_{B^*}^2 - M_B^2 + O\left(\frac{1}{M_B}\right)\ 
\end{array}
\label{eq:lambda2def}
\end{equation}
and $h$ is the field of the (static) heavy quark.

In order to avoid the renormalon ambiguity it is necessary to
introduce a hard ultraviolet cut-off (or subtraction scale), $\mu$, so
as to provide a well-defined boundary between long- and short-distance
contributions. In lattice calculations this occurs naturally; the
lattice spacing $a$ is such a cut-off ($\mu=a^{-1}$). In continuum
calculations, a hard cut-off has been used in
refs:\cite{bigipower,kapustinpower,melnikov}.

With a hard cut-off it is clear that the matrix elements of higher
dimensional operators diverge as powers of $\mu$ and in themselves do
not have a natural physical interpretation, e.g.
\begin{equation}
\langle f|O_1(\mu)|i\rangle\sim\mu^n\ .
\label{eq:o1diverge}\end{equation}
Since physical quantities cannot depend on the cut-off, the
term proportional to $\mu^n$ present in the matrix element in
eq.(\ref{eq:o1diverge})  must be cancelled in the full
prediction of ${\cal P}$. Specifically, this term is cancelled
by corresponding ones in the coefficient function $C_0$, which
now takes the form
\begin{equation}
C_0=\sum_{i=0}c_i\alpha_s^i + \frac{\mu^n}{m_b^n}\sum_{i=1}
d_i\alpha_s^i\ .
\label{eq:c0expand}\end{equation}
The second term on the right hand side of eq.~(\ref{eq:c0expand}) arises
from the matching of full QCD onto the hqet, when the operator $O_1$ is
included in the OPE. The important point is that the coefficient
function $C_0$ is evaluated perturbatively and hence, in practice, this
cancellation will only be achieved partially. In order to evaluate (or
even to estimate) the $O(\Lambda_{QCD}^n/m_b^n)$ corrections to ${\cal
P}$, we must evaluate $C_0$ to a sufficiently high order of perturbation
theory to reach the corresponding precision. We believe that this is not
the case in present calculations, and certainly has not been
demonstrated to be so~\footnote{In lattice calculations it may be
possible to calculate the coefficients $d_i$ to reasonably high orders
by using the Langevin stochastic formulation of lattice field
theory~\cite{direnzo}.}.

For calculations in heavy quark physics the above discussion implies
that there is no ``natural'' definition of parameters such as the
binding energy $\bar\Lambda$ (see eq.~(\ref{eq:ylambdabar})\,)
or the kinetic energy $\lambda_1$
\begin{equation}
\lambda_1(B)=\frac{1}{2m_B}\langle B|\,\bar h(i\vec D)^2 b\,|B
\rangle\ .
\label{eq:l1def}\end{equation}
Different definitions of these parameters, due to different
renormalization prescriptions for the operators, differ by terms of
$O(\Lambda_{QCD})$ and $O(\Lambda_{QCD}^2)$ respectively. It is
therefore of little use to compare values of these parameters, obtained
using different definitions (often the definitions are implicit and must
be inferred from the details of the calculations); nevertheless, this is
frequently done.

\section{Non-Leptonic Inclusive Decays}
\label{sec:nlinclusive}

In this section I discuss two very interesting problems in the
phenomenology of $B$-decays, that of lifetimes and the semileptonic
branching ratio. The discussion will use the formalism of Bigi \etal
(see ref.~\cite{bsuv2} and references therein), developed and used by
them and many other groups, in which inclusive quantities are expanded
in inverse powers of the mass of the heavy quark, e.g.
\begin{equation}
\Gamma(H_b) = \frac{G_F^2m_b^5|V_{cb}|^2}{192\pi^3}
\left\{c_3\left(1+ \frac{\lambda_1+3\lambda_2}{2m_b^2}\right)
+c_5\frac{\lambda_2}{m_b^2}+O\left(\frac{1}{m_b^3}\right)\right\}\ ,
\label{eq:widthexpansion}\end{equation}
where $\Gamma$ is the full or partial width of a beauty hadron $H_b$,
$c_{3,5}$ are coefficients which can be computed in perturbation theory
and $\lambda_{1,2}$ are the parameters introduced in
section~\ref{sec:power} above. Here I will not rediscuss the
cancellation of renormalon ambiguities present in $c_3$, $\lambda_1$ and
the quark mass; below we will consider ratios of physical quantities for
which the cancellation is more transparent. An important feature of the
general expression in eq.~(\ref{eq:widthexpansion}) is the absence of
terms of $O(1/m_b)$, which is a consequence of the absence of any
operators of dimension 4 which can appear in the corresponding
OPE~\cite{nooneoverm}.

\subsection{Beauty Lifetimes}
\label{subsec:lifetimes}

Using the expression in eq.~(\ref{eq:widthexpansion}) for the widths one
readily finds the following results for the ratios of lifetimes:
\begin{eqnarray}
\frac{\tau(B^-)}{\tau(B^0)} & = & 1 + O\left(\frac{1}{m_b^3}\right)
\label{eq:ratiomesons}\\
\frac{\tau(\Lambda_b)}{\tau(B^0)} & = &
1 + \frac{\mu_\pi^2(\Lambda_b)-\mu_\pi^2(B)}{2m_b^2} +
c_G \frac{\mu_G^2(\Lambda_b)-\mu_G^2(B)}{m_b^2}
+ O\left(\frac{1}{m_b^3}\right)\nonumber\\ 
& = & (0.98\pm0.01) + O\left(\frac{1}{m_b^3}\right)\ ,
\label{eq:ratiolambdab}\end{eqnarray}
where $\mu_\pi^2 = -\lambda_1$ and $\mu_G^2=3\lambda_2$.
In order to obtain the result in eq.(\ref{eq:ratiolambdab}), one needs
to know the difference of the kinetic energies of the $b$-quark in the
baryon and meson. To leading order in the heavy quark expansion we have:
\begin{equation}
\mu_\pi^2(\Lambda_b)-\mu_\pi^2(B) = -\frac{M_BM_D}{2}
\left(\frac{M_{\Lambda_b} - M_{\Lambda_c}}{M_B-M_D} - \frac{3}{4}
\frac{M_{B^*}- M_{D^*}}{M_B - M_D}-\frac{1}{4}\right)\ .
\label{eq:kediffs}\end{equation}
From equation~(\ref{eq:kediffs}), and using the recent measurement of
$m_{\Lambda_B}$ from CDF, one finds that the right hand side is very
small (less than about 0.01 GeV$^2$). The matrix elements of the
chromomagnetic operator are obtained from the mass difference of the
$B^*$- and $B$-mesons (see eq.(\ref{eq:lambda2def}) and from the
fact that the two valence quarks in the $\Lambda_b$ are in a spin-zero
state. The theoretical predictions in eqs.~(\ref{eq:ratiomesons}) and
(\ref{eq:ratiolambdab}) can be compared with the experimental measurements
\begin{equation}
\begin{array}{ccc}
\displaystyle
\frac{\tau(B^-)}{\tau(B^0)}  =  1.06 \pm 0.04 &\ \mbox{and}\ &
\displaystyle
\frac{\tau(\Lambda_b)}{\tau(B^0)}  =  0.79\pm 0.05\ .
\label{eq:ratiolambdabexp}\end{array}\end{equation}
The discrepancy between the theoretical and experimental results
for the ratio $\tau(\Lambda_b)/\tau(B^0)$ in
eqs.~(\ref{eq:ratiolambdab}) and (\ref{eq:ratiolambdabexp}) is
notable. It raise the question of whether the $O(1/m_b^3)$
contributions are surprisingly large, or whether there is a more
fundamental problem. I postpone consideration of the latter
possibility and start with a discussion of the $O(1/m_b^3)$ terms.

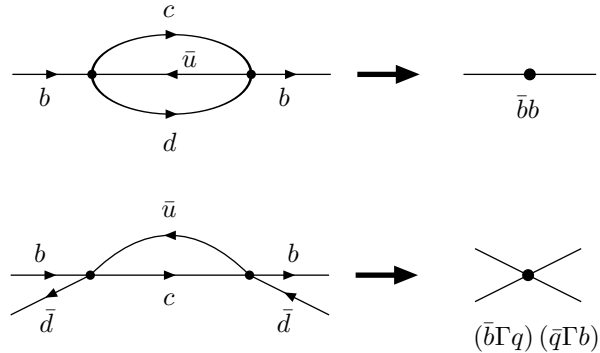
\begin{figure}[ht]
\begin{center}
\begin{picture}(300,50)(-30,15)
\ArrowLine(10,41)(40,41)\ArrowLine(100,41)(40,41)\ArrowLine(100,41)(130,41)
\Oval(70,41)(15,30)(0)
\GCirc(40,41){1.5}{0}\GCirc(100,41){1.5}{0}
\ArrowLine(69,56)(71,56)\ArrowLine(69,26)(71,26)
\Text(20,36)[tl]{$b$}\Text(115,36)[tr]{$b$}
\Text(67,63)[bl]{$c$}\Text(67,19)[tl]{$d$}\Text(80,44)[br]{$\bar u$}
\SetWidth{2}\Line(140,41)(158,41)\ArrowLine(158,41)(160,41)
\SetWidth{0.5}\Line(180,41)(230,41)\GCirc(205,41){2}{0}
\Text(205,33)[t]{$\bar bb$}
\end{picture}
\begin{picture}(300,50)(260,40)
\ArrowLine(300,41)(330,41)\ArrowLine(330,41)(300,26)
\ArrowLine(330,41)(390,41)\Curve{(330,41)(360,56)(390,41)}
\ArrowLine(390,41)(420,41)\ArrowLine(420,26)(390,41)
\ArrowLine(361,56)(359,56)
\GCirc(330,41){1.5}{0}\GCirc(390,41){1.5}{0}
\SetWidth{2}\Line(430,41)(448,41)\ArrowLine(448,41)(450,41)
\SetWidth{0.5}\Line(475,51)(495,41)\Line(495,41)(515,51)
\Line(495,41)(475,31)\Line(515,31)(495,41)
\GCirc(495,41){2}{0}
\Text(310,46)[bl]{$b$}\Text(312,28)[tl]{$\bar d$}
\Text(361,63)[b]{$\bar u$}\Text(361,34)[t]{$c$}
\Text(410,46)[br]{$b$}\Text(407,28)[tr]{$\bar d$}
\Text(500,23)[t]{$(\bar b\Gamma q)\,(\bar q\Gamma b)$}
\end{picture}
\vspace{0.4in}
\caption{Examples of diagrams whose imaginary parts contribute to
the total rates for the decays of beauty hadrons (left-hand sides)
and the operators they correspond to in the Operator Product Expansions.
$\Gamma$ represents a Dirac matrix.\label{fig:diagrams}}
\end{center}\end{figure}

At first sight it seems strange to consider the $1/m_b^3$ corrections to
be a potential source of large corrections, when the $O(1/m_b^2)$ terms
are only about 2\%. However, it is only at this order that the
``spectator'' quark contributes, and so these contributions lead
directly to differences in lifetimes for hadrons with different light
quark constituents (consider for example the lower diagram in
Fig.~\ref{fig:diagrams}, for which, using the short-distance expansion,
one obtains operators of dimension 6).  Moreover, the coefficient
functions of these operators are relatively large, which may be
attributed to the fact that the lower diagram in Fig.~\ref{fig:diagrams}
is a one-loop graph, whereas the corresponding diagrams for the leading
contributions are two-loop graphs (see, for example, the upper diagram
of Fig.~\ref{fig:diagrams}). The corresponding phase-space enhancement
factor is 16$\pi^2$ or so. We will therefore only consider the
contributions from the corresponding four-quark operators, neglecting
other $O(1/m_b^3)$ corrections which do not have the phase space
enhancement~\cite{ns}.  For each light-quark flavour $q$, there are four
of these~\footnote{I use the notation of ref.~\cite{ns}.}:
\begin{eqnarray}
O_1\equiv \bar b\gamma_\mu(1-\gamma^5)q\, \bar q\gamma^\mu(1-\gamma^5)b
\hspace{0.32in} &;&
O_2\equiv \bar b (1+\gamma^5)q\, \bar q(1+\gamma^5)b
\label{eq:odefs}\\
T_1\equiv \bar b\gamma_\mu(1-\gamma^5)T^aq \,\bar q\gamma^\mu(1-\gamma^5)T^a b
&;&  
T_2\equiv \bar b (1+\gamma^5)T^a q \,\bar q (1+\gamma^5) T^a b
\end{eqnarray}
where $T^a$ are the generators of colour $SU(3)$. Thus we need to
evaluate the matrix elements of these four operators.

For mesons, following ref.~\cite{ns}, I introduce the parametrization
\begin{equation}
\begin{array}{ccc}
\langle B| O_i|B\rangle|_{\mu = m_b} \equiv
B_if_B^2M_B^2 & ; &
\langle B| T_i|B\rangle|_{\mu = m_b}  \equiv
\epsilon_if_B^2M_B^2\ ,
\end{array}
\label{eq:biepsilonidef}\end{equation} 
where $\mu$ is the renormalization scale. We have chosen to use $m_b$ as
the renormalization scale. Bigi et al.~\cite{bsuv2}  prefer to use a
typical hadronic scale, and estimate the matrix elements using a
factorization hypothesis at this low scale. Operators renormalized at
different scales can be related using renormalization group equations in
the hqet (sometimes called hybrid renormalization~\cite{hybrid}). For
example, if we assume that factorization holds at a low scale $\mu$ such
that $\alpha_s(\mu^2)=1/2$, then, using the (leading order)
renormalization group equations, one finds $B_1=B_2=1.01$ and
$\epsilon_1=\epsilon_2=-0.05$~\footnote{By factorization we mean that if
the $B_i$'s and $\epsilon_i$'s had been defined at this scale (instead
of $m_b$) they would have been 1 and 0 respectively.}\,. In the limit of
a large number of colours $N_c$, $B_i=O(N_c^0)$ whereas
$\epsilon_i=O(1/N_c)$.

For the $\Lambda_b$, heavy quark symmetry implies that
\begin{equation}
\begin{array}{ccc}
\displaystyle
\langle\Lambda_b|O_2|\Lambda_b\rangle=
-\frac{1}{2}\langle\Lambda_b|O_1|\Lambda_b\rangle
&\ \mbox{and}\ &\displaystyle
\langle\Lambda_b|T_2|\Lambda_b\rangle=
-\frac{1}{2}\langle\Lambda_b|T_1|\Lambda_b\rangle\ ,
\label{eq:lbhqs}\end{array}\end{equation}
so that there are only two parameters. It is convenient to replace
the operator $T_1$, by $\tilde O_1$ defined by
\begin{equation}
\tilde O_1 \equiv \bar b^i\gamma_\mu(1-\gamma^5) q^j\ 
\bar q^j\gamma^\mu(1-\gamma^5) b^i\ ,
\end{equation}
where $i,j$ are colour labels,
and to express physical quantities in terms of the two parameters
$\tilde B$ and $r$ defined by
\begin{eqnarray}
\langle\Lambda_b|\tilde O_1|\Lambda_b\rangle_{\mu=m_b} & \equiv &
-\tilde B \langle\Lambda_b|O_1|\Lambda_b\rangle_{\mu=m_b}
\label{eq:btildedef}\\
\frac{1}{2M_{\Lambda_b}}\langle\Lambda_b|O_1|\Lambda_b\rangle_{\mu=m_b}
& \equiv & -\frac{f_B^2M_B}{48}\,r
\label{eq:rdef}\end{eqnarray}
We do not know the values of these parameters. In quark models $\tilde B=1$,
and $r=0.2$--$0.5$. Using experimental values of the hyperfine splittings and
quark models, it
has been suggested that $r$ may be larger~\cite{rosner}, e.g.
\begin{equation}
\begin{array}{ll}
r\simeq\frac{4}{3}\frac{M_{\Sigma_c^*}^2 - M_{\Sigma_c}^2}
{M_{D^*}^2 - M_{D}^2} = 0.9 \pm 0.1\label{eq:rc}\ \ 
&
\ \ r\simeq\frac{4}{3}\frac{M_{\Sigma_b^*}^2 - M_{\Sigma_b}^2}
{M_{B^*}^2 - M_{B}^2}  =  1.8 \pm 0.5\label{eq:rb}\ .
\end{array}\end{equation}

The lifetime ratios can now be written in terms of the six parameters
$B_{1,2},\epsilon_{1,2},\tilde B$ and $r$ (as well as $f_B$):
\begin{eqnarray}
\frac{\tau(B^-)}{\tau(B^0)} & = & 1 + \left(\frac{f_B}{200\mev}\right)^2
\left\{0.02B_1 + 0.00 B_2 -0.70 \epsilon_1 + 0.20\epsilon_2\right\}
\label{eq:mesonsresult}\\
\frac{\Lambda_b}{\tau(B^0)} & = & 0.98 + \left(\frac{f_B}{200\mev}\right)^2
\{-0.00B_1 + 0.00 B_2 -0.17 \epsilon_1 + 0.20\epsilon_2\nonumber\\ 
& & \hspace{0.6in} 
+ (-0.01 -0.02\tilde B)\,r\}\label{eq:baryonsresult}\ ,
\end{eqnarray}
where the effective weak Lagrangian has been renormalized at $\mu=m_b$.
The central question is whether it is possible, with ``reasonable" values
of the parameters, to obtain agreement with the experimental numbers in
eq.~(\ref{eq:ratiolambdabexp}). At this
stage in our knowledge, the answer depends somewhat on what is meant
by reasonable. For example, Neubert, guided by the arguments
outlined above,  has considered these ratios by
varying the parameters in the following ranges~\cite{mnhonolulu}:
\begin{equation}
B_i,\tilde B\in\left[\frac{2}{3},\frac{4}{3}\right]\,;\ \
\epsilon_i\in\left[-\frac{1}{3},\frac{1}{3}\right]\, ;\ \ r\in[0.25, 2.5]\,;
\ \ \left(\frac{f_B}{200\mev}\right)^2\in[0.8,1.2]\ .
\label{eq:ranges}\end{equation}
He concludes that, within these ranges, it is just possible to obtain
agreement at the two standard deviation level for large values of $r$
($r\ge 1.2$) and negative values of $\epsilon_2$. Lattice studies of the
corresponding matrix elements are underway; a recent  QCD sum-rule
calculation has found a small value of $r$, 
$r\simeq 0.1$--$0.3$~\cite{colangelo}.

If the lattice calculations confirm that the parameter $r$ is small, or
find that the other parameters are not in the appropriate ranges, then
we have a breakdown of our understanding. If no explanation can be found
within the standard formulation, then we will be forced to take
seriously the possible breakdown of local duality. This is beginning to
be studied in toy field theories~\cite{blok,nardulli}.

\subsection{The ``Baffling" Semileptonic Branching Ratio}
\label{subsec:bsl}

\begin{figure}[t]
\hbox to\hsize{\hss\vbox{\offinterlineskip
\unit=0.4\hsize
\epsfxsize=\unit\epsffile{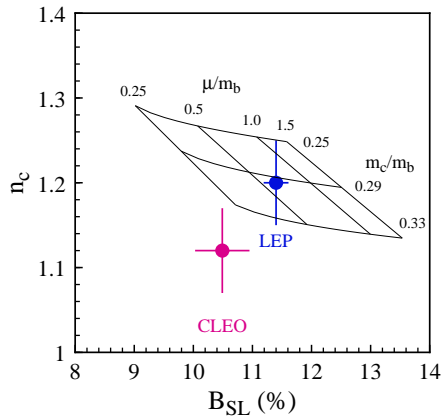}
}\hss}
\caption[]{Theoretical Prediction 
of the semileptonic branching ratio and charm counting. The
data points are the experimental results from high-energy
(LEP) and low energy (i.e. at the $\Upsilon(4S)$ from CLEO)
experiments.}
\label{fig:bsl}
\end{figure}

This was the name given by Blok \etal~\cite{baffling} to the
observation that the experimental value of the semileptonic branching
ratio 
\begin{equation}
B_{SL} = \frac{\Gamma(B\to Xe\bar\nu)}{\sum_{l}\Gamma(B\to Xl\bar\nu)
+ \Gamma_{had} + \Gamma_{rare}}
\label{eq:bsldef}\end{equation}
appeared to be lower than expected theoretically. In
eq.~(\ref{eq:bsldef}) the sum is over the three species of lepton, and
$\Gamma_{had}$ and $\Gamma_{rare}$ are the widths of the hadronic and
rare decays respectively. Bigi \etal\ concluded that a branching ratio
of less than 12.5\% cannot be accommodated by theory~\cite{baffling}.
Since then Bagan \etal\  have completed the calculation of the
$O(\alpha_s)$ corrections, and in particular of the $b\to c\bar cs$
component (including the effects of the mass of the charm
quark)~\cite{bagan}; these have the effect of decreasing $B_{SL}$. 
With M.~Neubert, we used this input to reevaluate the branching
ratio and charm counting ($n_c$, the average number of charmed particles
per $B$-decay)~\cite{ns} finding, e.g.
\begin{equation}
\begin{array}{ll}
B_{SL} = 12.0\pm 1.0\%\ (\mu=m_b)\ \ & 
\ \ n_c = 1.20 \mp 0.06\ \ (\mu=m_b)\label{eq:bslncmb}\\ 
B_{SL} = 10.9\pm 1.0\%\ (\mu=m_b/2)\ \ &
\ \ n_c = 1.21 \mp 0.06\ \ (\mu=m_b/2)\label{eq:bslnchalfmb}\ .
\end{array}\end{equation}
$\mu$ is the renormalization scale and the dependence on this scale is a
reflection of our ignorance of higher order perturbative corrections.
The experimental situation is somewhat confused, see Fig.~\ref{fig:bsl}.
In his compilation at the ICHEP conference last year~\cite{richman}, Richman
found that the semileptonic branching ratio obtained from $B$-mesons
from the $\Upsilon(4S)$  is~\footnote{Note that the rapporteur at the
1997 EPS conference argued that the branching ratio had been
overestimated by the LEP collaboration~\cite{feindt}.}:
\begin{equation}
B_{SL}(B)=(10.23\pm0.39)\%\ ,
\end{equation}
whereas that from LEP is:
\begin{equation}
B_{SL}(b)=(10.95\pm0.32)\%\ .
\end{equation}
The label $b$ for the LEP measurement indicates that 
the decays from beauty hadrons other than the $B$-meson are included.
Using the measured fractions
of the different hadrons and their lifetimes, and assuming that
the semileptonic widths of all the beauty hadrons are the same,
one finds:
\begin{equation}
B_{SL}(b)=(10.95\pm 0.32)\% \Rightarrow B_{SL}(B)=(11.23\pm0.34)\%\ , 
\end{equation}
amplifying the discrepancy. It is very difficult to understand such a
discrepancy theoretically, since the theoretical calculation only
involves $\Gamma_{SL}$ (and not $\Gamma_{had}$ for which the
uncertainties are much larger). In view of the experimental
discrepancy, I consider the problem of the lifetime ratio
$\tau(\Lambda_b)/\tau(B^0)$, described in
subsection~\ref{subsec:lifetimes} above, to be the more significant one.

\section{Exclusive Nonleptonic Decays}
\label{sec:nlexc}

In this section I consider two-body nonleptonic decays of $B$-mesons,
for which a large amount of data is becoming available, particularly
from the CLEO collaboration~\cite{drell}. This is an exciting new
field of investigation, which will undoubtedly teach us much about
subtle aspects of the standard model.  Unfortunately, at our present
level of understanding we are not able to compute the amplitudes from
first principles, and are forced to make assumptions about the
non-perturbative QCD effects; frequently these assumptions concern
factorization.  {\em These assumptions may well be wrong}. Thus the
analyses are limited to a semi-quantitative level.  In this talk I
will briefly describe some recent attempts to understand nonleptonic
exclusive decays; at this stage it is not possible to endorse these
approaches with any confidence.

\subsection{Generalized Factorization Hypothesis}

Neubert and Stech suggest an approach based on keeping the leading
order terms in the limit of a large number of colours
($N_c$)~\cite{stech}.  Consider $B\to D\pi$ decays, for which the effective
Hamiltonian is given by:
\begin{equation}
{\cal H}_{eff} = \frac{G_F}{\sqrt{2}}
V_{cb}V_{ud}^*\left\{c_1(\mu)\,(\bar du)\,(\bar cb) +
c_2(\mu)\,(\bar cu)\,(\bar db)\right\}\ ,
\label{eq:heffnl}\end{equation}
where, at the renormalization scale $\mu=m_b$, $c_1(m_b)=1.13$ and 
$c_2(m_b)=-0.29$, and the $V-A$ structure of the current is implied,
e.g. $(\bar du)\,(\bar cb)=(\bar d\gamma^\mu(1-\gamma^5)u)\,
(\bar c\gamma_\mu(1-\gamma^5)b)$.

For the class-1 decay $B^0\to D^+\pi^-$, using colour and spinor
Fierz identities we can write:
\begin{equation}
A_{B^0\to D^+\pi^-} = \left(c_1 + \frac{c_2}{N_c}\right)
\langle D^+\pi^-\,|\,(\bar d u)\,(\bar c b)|B^0\rangle
+2c_2\,\langle D^+\pi^-\,|\,
(\bar d T^a u)\,(\bar c T^a b)|B^0\rangle\ .
\label{eq:ab0todplus}\end{equation}
Following ref.~\cite{stech} we write 
\begin{equation}
{\cal F}_{(BD)\pi}\equiv \langle\pi^-\,|\,(\bar du)\,|0\rangle
\langle D^+\,|\,(\bar cb)\,|\,B^0\rangle 
\label{eq:calfdef}\end{equation}
and
\begin{equation}\begin{array}{ccc}
\displaystyle
A_{B^0\to D^+\pi^-} \equiv a_1{\cal F}_{(BD)\pi}, & \ \mbox{with}\ &
\displaystyle
a_1= \left( c_1 +
\frac{c_2}{N_c}\right)\left[1+\epsilon^{(BD)\pi}_1\right]
+c_2\epsilon^{(BD)\pi}_8\ .
\end{array}\label{eq:epsdefs}\end{equation}
So far this is only a parametrization of the amplitude. The two
factors on the right-hand side of eq.~(\ref{eq:calfdef}) are given in
terms of the decay constant $f_\pi$ and the form-factors for
semileptonic $B\to D$ decays respectively. The naive factorization
hypothesis would imply that $a_1 = c_1 + c_2/N_c$. Neubert and Stech
argue that the large-$N_c$ expansion may be a better guide than
factorization, in which case we have $a_1 = c_1 + O(1/N_c^2)$.

For the class-2 process $B^0\to D^0\pi^0$, the relations corresponding to 
eqs.~(\ref{eq:ab0todplus}) to (\ref{eq:epsdefs}) are:
\begin{eqnarray}
A_{B^0\to D^0\pi^0} & = & \left(c_2 + \frac{c_1}{N_c}\right)
\langle D^0\pi^0\,|\,(\bar c u)\,(\bar d b)|B^0\rangle
+2c_1\langle D^0\pi^0\,|\,
(\bar c T^a u)\,(\bar d T^a b)|B^0\rangle
\label{eq:ab0todzero}\\ 
{\cal F}_{(B\pi)D}& \equiv & \langle\,D^0\,|\,(\bar cu)\,|0\rangle
\langle \pi^0\,|\,(\bar db)\,|\,B^0\rangle
\label{eq:calf2def}\\ 
A_{B^0\to D^+\pi^-}& \equiv& a_2{\cal F}_{(B\pi)D}
\label{eq:a2def}\\ 
a_2 & = &\left( c_2 +
\frac{c_1}{N_c}\right)\left[1+\epsilon^{(B\pi)D}_1\right]
+c_1\epsilon^{(B\pi)D}_8\ .
\label{eq:epsdefs2}\end{eqnarray}
In this case naive factorization would imply that
$a_2 = c_2 + c_1/N_c$, whereas using the
large $N_C$ expansion $a_2 = c_2 + c_1(1/N_c + \epsilon^{(B\pi)D}_8)
+O(1/N_c^3)$, (the leading terms in $a_2$ are all of $O(1/N_c)$).
This approximation preserves the correct renormalization group behaviour
(up to corrections of $O(1/N_c^3)$).

Neubert and Stech propose a ``generalized'' factorization hypothesis for
processes in which a large amount of energy is released, based on the
large $N_C$ expansion described above, and on the concept of colour
transparency applied also to class-2 decays~\cite{transparency} (for
more formal arguments see also \cite{dugan})~\footnote{Neubert and Stech
also apply factorization assumptions to processes in which the energy of
the outgoing particles are not necessarily large~\cite{stech}.}
\begin{equation}
a_1=c_1(m_b)\ \ \mbox{and}\ \ \ a_2 = c_2(m_b) + \zeta c_1(m_b)\ ,
\label{eq:genfact}\end{equation}
where $\zeta$ is a process independent parameter (for two-body decays). 
They study a wide
variety of processes and conclude that $a_{1,2}$ are process
independent within the available precision:
\begin{equation}\begin{array}{ccc}
a_1  =  1.10 \pm 0.07\pm 0.17 &\ \mbox{and}\ \ 
a_2 =  0.21 \pm 0.01\pm 0.04\label{eq:a1a2result} \ .
\end{array}\end{equation}

It is now important, not only to extend the theoretical and experimental
studies to improve the precision and determine the range of validity of
the hypothesis, but above all to try and understand the theoretical
foundations (if any) for it.

\subsection{$B$-Decays to Two Light Hadrons} 

The recent data from the CLEO collaboration on $B$-decays into two
light mesons has stimulated many theoretical papers (see
e.g. refs.~\cite{ag,charming} and for decays (inclusive as well as
exclusive) with an $\eta^\prime$ in the final state in particular see
refs.~\cite{btoll}).  Many of the decays have suppressed tree level
contributions, so that loop effects, which are sensitive to the
presence of new physics, are important. As always, the principal
difficulty in drawing quantitative conclusions from the experimental
data is our inability to control the non-perturbative QCD effects.
For example, control of the penguin contribution to the decay
$B\to\pi^+\pi^-$, which may well be significant, is needed in the
forthcoming studies of $CP$-violation and in the determination of the
angle $\alpha$ of the unitarity triangle.

Among the recent phenomenological studies of the CLEO data is a standard
analysis by Ali and Greub~\cite{ag} who include the next-to-leading
order coefficient functions, and estimates of the hadronic matrix
elements based on factorization.  They find that all the $B\to K\pi$ and
$\pi\pi$ rates or upper limits can be accommodated, within this picture.
Their estimates of the effects of charm quarks in penguin diagrams are
based on perturbation theory. This has been criticized by Ciuchini
\etal\ who stress that the relevant regions of phase space are close to
the charm threshold and hence intrinsically
non-perturbative~\cite{charming}. Their book-keeping of penguin effects
within the OPE is briefly outlined in the next subsection. Concerning
the phenomenological conclusions, Ciuchini et al. question whether the
consistency of theoretical predictions and experimental measurements or
bounds found in ref.~\cite{ag} can be achieved for all processes with a
single set of non-perturbative parameters. For example, Ciuchini et al.
find it difficult to satisfy simultaneously the experimental branching
ratio for the process $B\to\eta'K^0$ and the bound on the branching
ratio for $B_d\to\pi^+\pi^-$. Such debates in the community, on what is a
very new field of study, will lead to considerable insights into subtle
features of the Standard Model.

\subsection{Charming Penguins}
\label{subsec:charming}

The analysis of non-leptonic $b$-decays generally starts with a
classification of the relevant operators and a calculation of their
coefficient functions in perturbation theory. For example:
\begin{equation}
\begin{array}{lcl}
Q_1^u = (\bar bd)_{V-A}(\bar uu)_{V-A} &\mbox{\hspace{0.1in}}
& Q_1^c = (\bar bd)_{V-A}(\bar cc)_{V-A} \label{eq:q1def}\\ 
Q_2^u = (\bar bu)_{V-A}(\bar ud)_{V-A} &\mbox{\hspace{0.1in}}
& Q_2^c = (\bar bc)_{V-A}(\bar cd)_{V-A} \label{eq:q2def}\ ,
\end{array}
\end{equation}
plus QCD and electroweak penguin operators. The coefficients of
$Q_{1,2}^{u,c}$ are of $O(1)$ whereas those of the remaining operators
are of $O(\alpha_s)$ and are generally small. 

Ciuchini \etal~\cite{charming} have recently emphasized the point,
previously made by Buras and Fleischer~\cite{bfcharming} that the matrix
elements of the operators $Q_{1,2}^{u,c}$ have penguin-like contractions
(when the up or charm quark fields are Wick contracted). For the case of
the charm quark, the effects of these ``charming'' penguins are enhanced
in decays where the emission diagrams are Cabibbo suppressed relative to
the penguin diagram. Although the precise definition of charming
penguins (i.e. the  separation of penguin effects between the explicit
contributions of the further operators $Q_{3,\cdots}$ and those in the
matrix elements of the operators $Q_{1,2}$) is a matter of convention,
they do contain non-perturbative QCD effects which must be included in
the calculations of decay amplitudes. From their recent analysis,
Ciuchini \etal\  conclude that large contributions are likely from
charming penguins, e.g. that it is necessary to include them in order to
obtain branching ratios of order 1$\times 10^{-5}$ for the decays
$B^+\to K^0\pi^+$ and $B_d\to K^+\pi^-$ as recently found by the CLEO
collaboration~\cite{cleobtopik}.  By using recent results from CLEO,
Ciuchini \etal\  constrain many of the relevant hadronic matrix elements
and are able to make predictions for a large number of processes which
have not yet been observed~\cite{charming}. Some of these processes,
including $B\to\rho K$, are predicted to have branching ratios close to
the current experimental bounds, and will provide an excellent testing
ground for our understanding of penguin effects.

\subsection{Lattice Calculations}
\label{subsec:nllattice}

The amplitudes for non-leptonic decays of hadrons ($H\to h_1h_2$) are
difficult to evaluate in lattice simulations in
principle. Maiani and Testa pointed out that in lattice simulations one
obtains the following average of matrix elements~\cite{mate}:
\begin{equation}
\frac{1}{2}\left(\, _{in}\langle h_1 h_2\,|{\cal H}_{eff}\,|H\rangle
+ _{out}\langle h_1 h_2\,|{\cal H}_{eff}\,|H\rangle\,\right)\ ,
\label{eq:inout}\end{equation}
and hence no information about the phase of the final state interactions.
They also show that from the large time behaviour of the correlation
functions one can extract the unphysical matrix element:
\begin{equation}
\langle h_1(\vec p_{h_1}=\vec 0) h_2(\vec p_{h_2}=\vec 0) |{\cal
H}_{eff}|H\rangle\ .
\label{eq:rest}\end{equation}
Together with chiral perturbation theory, this may be useful to obtain
good estimates of the amplitudes for kaon decays, but it is not very
useful for $B$-decays.

The Maiani-Testa theorem~\cite{mate} implies that it is not possible
to obtain the phase of the final state interactions without some
assumptions about the amplitudes. The importance of developing
reliable quantitative techniques for the evaluation of
non-perturbative QCD effects in non-leptonic decays cannot be
overstated, and so attempts to introduce ``reasonable'' assumptions to
enable calculations to be performed (and compared with experimental
data) are needed urgently.  Ciuchini et
al.~\cite{ciuchini,silvestrini} have recently shown that by making a
``smoothness'' hypothesis about the decay amplitudes it is possible to
extract information about the phase of two-body non-leptonic
amplitudes. Studies to see whether their proposals are practicable and
consistent are currently beginning.

\section{Conclusions}
\label{sec:concs}

During recent years, through the combined work of experimentalists and
theorists, there has been enormous progress in our understanding of
heavy quark physics. In this talk I have reviewed some of this work,
and underlined a few of the areas in which further progress is urgently
needed in developing control of the non-perturbative QCD effects. 
Among the main points were:
\begin{itemize}
\item The lattice community will continue to refine the computations
of a wide variety of physical quantities (e.g. leptonic decay constants,
$B$-parameter of $B$-$\bar B$ mixing, form factors of semileptonic decays,
the parameters of the hqet), reducing the systematic errors, in 
particular those due to the use of the quenched approximation.
\item $V_{ub}$ can be determined by studying the hadronic invariant
mass spectrum in inclusive semileptonic decays. The authors of 
ref.~\cite{flw} estimate that this will lead to an error of about
10\% in $V_{ub}$ whereas those of ref.~\cite{dikeman} that the error
will not exceed 10--20\%.
\item We need to understand whether the discrepancy of the theoretical
prediction for the ratio $\tau(\Lambda_b)/\tau(B)$ with the corresponding
experimental measurement is due to some effect which can be controlled
(such as a matrix elements which is larger than currently expected) or
to violations of local duality.
\item We need progress in understanding whether power corrections
to hard scattering and decay processes can be controlled numerically.
\item New theoretical ideas are urgently required to interpret
quantitatively the wealth of data on non-leptonic exclusive decays from
current and future experiments. In particular progress in this area is
needed for studies of CP-violation at the forthcoming $b$-factories, and
in the attempts to fix the angles of the uniitarity triangle.
\end{itemize}

\section*{Acknowledgments} 

I warmly thank P.~Ball, C.~Bernard, I.~Bigi, V.~Braun, A.~Buras, P.~Drell,
J.~Flynn, S.~Hashimoto, G.~Martinelli, T.~Mannel, M.~Neubert, T.~Onogi
and Y.~Nir for discussions and help in clarifying important questions.
 
I acknowledge support from the Particle Physics and Astronomy Research
Council, UK, through grants GR/K55738 and GR/L22744.
 
\section*{References}
\def\warsaw{ICHEP96, 28th Int. Conf. on High Energy Physics, Warsaw,
Poland, 25--31 July 1996, edited by Z. Ajduk and A.K. Wroblewski,
World Scientific, Singapore (1997)}
\def\edlat{Lattice 97, 15th Int. Symp. on Lattice Field Theory,
Edinburgh, Scotland, 1997}
\def\nllat{Lattice 92, 10th Int. Symp. on Lattice Field Theory,
Amsterdam, Netherlands, 1992}
\def\stlolat{Lattice 96, 14th Int. Symp. on Lattice Field Theory, St
Louis, USA, 1996}
\def\ozlat{Lattice 95, 13th Int. Symp. on Lattice Field Theory,
Melbourne, Australia, 1995}
\def\prd#1{Phys. Rev. D {\bf #1}}
\def\prl#1{Phys. Rev. Lett. {\bf #1}}
\def\plb#1{Phys. Lett. B {\bf #1}}
\def\npb#1{Nucl. Phys. B {\bf #1}}
\def\npbps#1{Nucl. Phys. B (Proc. Suppl.) {\bf #1}}
\def\npaps#1{Nucl. Phys. A (Proc. Suppl.) {\bf #1}}
\def\zpc#1{Z. Phys C {\bf #1}}
\def\nima#1{Nucl. Instrum. Meth. A {\bf #1}}
\def\cmp#1{Commun. Math. Phys. {\bf #1}}
\def\physrep#1{Phys. Rep. {\bf #1}}

\end{document}